\documentclass[aps,prd,showpacs,nofootinbib,reprint]{revtex4-2}

\usepackage{booktabs}
\usepackage{stmaryrd}
\usepackage{mathrsfs}
\usepackage{amsmath}
\usepackage{latexsym}
\usepackage{amsfonts}
\usepackage{graphicx}
\usepackage{subfigure}
\usepackage{comment}
\usepackage{color}
\usepackage{hyperref}
\hypersetup{
  colorlinks=true,
  linkcolor=red,
  citecolor=blue,
}

\begin{document}
\title{Constant-roll inflation with non-minimally derivative coupling}

\author{Jie Liu}
\email{ liujie@hainnu.edu.cn} 
\affiliation{School of Physics and Electronic Engineering, Hainan Normal University, Haikou 571158, China}
\author{Yungui Gong}
\email{gongyungui@nbu.edu.cn} 
\affiliation{Department of Physics, School of Physical Science and Technology, Ningbo University, Ningbo, Zhejiang 315211, China}
\affiliation{School of Physics, Huazhong University of Science and Technology, Wuhan, Hubei
430074, China}
\author{Zhu Yi}
\email{Corresponding author. yz@bnu.edu.cn}
\affiliation{Faculty of Arts and Sciences, Beijing Normal University, Zhuhai 519087, China}
\affiliation{Advanced Institute of Natural Sciences, Beijing Normal University, Zhuhai 519087, China}
\begin{abstract}
We investigate the constant-roll inflation with non-minimally kinetic coupling to the Einstein tensor. 
With the slow-roll parameter $\eta_\phi = -\ddot{\phi}/(H\dot{\phi})$ being a constant, 
we calculate the power spectra for scalar and tensor perturbations, 
and derive the expressions for the scalar spectral tilt $n_s$,
the tensor spectral tilt $n_T$, and the tensor-to-scalar ratio $r$.
We find that the expressions for $n_s$ are different with different ordering of taking the derivative of the scalar power spectrum with respect to the scale $k$ and the horizon crossing condition $c_sk=aH$ in the constant-roll inflation,
the consistency relation $r=-8n_T$ does not hold if $|\eta_\phi|$ is not small,
and the duality of the tensor-to-scalar ratio between the slow-roll inflation and ultra-slow-roll inflation does not exist in inflationary models with non-minimally derivative coupling.
The result offers a fresh perspective on the understanding of the inflationary models with non-minimally derivative coupling and is helpful for the production of scalar induced gravitational waves in the framework of ultra-slow-roll inflation with non-minimally derivative coupling.
\end{abstract}

\maketitle
\section{Introduction}
Inflation, characterized by an exponential expansion of the Universe during its early stages, was proposed to address issues in standard Big Bang cosmology, including the horizon, flatness, and monopole problems \cite{Guth:1980zm,Linde:1981mu,Albrecht:1982wi,Starobinsky:1980te,Guth:1982ec}. 
This expansion is typically driven by a scalar field known as the inflaton, where the simplest model is the canonical inflaton field with a flat potential.  
Utilizing the Higgs field, the only detected scalar field in the Standard Model of particle physics, as the inflaton leads to a conflict with observations of the cosmic microwave background (CMB) radiation due to the large prediction of the tensor-to-scalar ratio $r$. 
To overcome it, inflation models with the field non-minimally coupled to gravity have been proposed  \cite{Bezrukov:2007ep,Germani:2010gm, Yi:2018gse}.  
One notable model is the new Higgs inflation model with the kinetic term non-minimally derivative coupling to the Einstein tensor,  represented by  $G^{\mu\nu}\partial_\mu\phi\partial_\nu\phi/M^2$ \cite{Germani:2010gm,Germani:2014hqa}.  
Under the general relativity (GR) limit $F=H^2/M^2 =0$, this model becomes the canonical situation conflicting with the observations, 
while under the high friction limit $F\gg1$, the tensor-to-scalar ratio is $1/3$ of that in the canonical Higgs inflation, aligning with the observational data \cite{Germani:2014hqa,Yang:2015pga}. 
Furthermore, the effective self-coupling $\lambda$ of the Higgs boson in this model can be lowered to the order of unity without introducing a new degree of freedom  \cite{Germani:2010gm,Germani:2014hqa}.  For more works about the inflation model with  non-minimally derivative coupling, see  Refs. \cite{Germani:2010ux,Minamitsuji:2013ura,Sadjadi:2012zp,Zhu:2015lry,Yang:2015zgh,Gong:2017kim,Fu:2019ttf, Granda:2019wip,Oliveros:2019xef,Gialamas:2020vto, Heydari:2021gea,Goodarzi:2022wli,Oikonomou:2024tjf,Gialamas:2024jeb,Hell:2024xbv}. 

Apart from the slow-roll inflation scenario, the constant-roll inflation scenario, where one slow-roll parameter remains constant and need not be small, can also yield predictions consistent with observational data \cite{Martin:2012pe,Motohashi:2014ppa,Yi:2017mxs}. 
The characteristics of primordial perturbations in constant-roll inflation can differ significantly from those in slow-roll inflation. For instance, in the canonical constant-roll inflation with a constant $\eta_\phi = -\ddot{\phi}/(H\dot{\phi})$ and $\eta_\phi>3/2$, scalar perturbations outside the horizon continue to evolve  \cite{Motohashi:2014ppa}. Additionally, the power spectrum of scalar perturbations can exhibit sharp enhancements at small scales in ultra-slow-roll inflation with $\eta_\phi \approx 3$, potentially leading to the formation of primordial black holes and inducing secondary gravitational waves \cite{Germani:2017bcs,Gong:2017qlj,Ezquiaga:2017fvi, Ballesteros:2017fsr,Drees:2019xpp,Lin:2020goi, Lin:2021vwc, Yi:2020cut,Yi:2020kmq,Yi:2021lxc,Yi:2022anu,Zhang:2021rqs,Wang:2024euw}. 
Despite the differences in the evolution of perturbations between ultra-slow-roll inflation  $\eta_\phi =3-\alpha$  and slow-roll inflation  $\eta_\phi =\alpha$  in the canonical inflation model \cite{Gao:2019sbz}, if we neglect the contribution from $\epsilon_H$ to the scalar spectral tilt $n_s$, the predictions for the scalar spectral tilt $n_s$ and tensor-to-scalar ratio $r$ are the same \cite{Tzirakis:2007bf,Morse:2018kda}. This indicates the existence of a duality in constant-roll inflation models with large and small $\eta_\phi$. 
For more about the constant-roll inflation, please refer to Refs. \cite{Motohashi:2017vdc, Oikonomou:2017bjx, Odintsov:2017qpp,Nojiri:2017qvx, Dimopoulos:2017ged,Ito:2017bnn,Karam:2017rpw,Fei:2017fub,Cicciarella:2017nls,Anguelova:2017djf,Gao:2018tdb,Gao:2018cpp,Mohammadi:2018wfk, Pattison:2018bct,Lin:2019fcz,Fei:2020jab,Antoniadis:2020dfq,Guerrero:2020lng,Shokri:2021iqp,Fei:2022gfp,Stojanovic:2023dni,Huang:2024xqk}. 

In this paper, we research the constant-roll inflation model with non-minimally derivative coupling.   
We calculate the power spectrum of scalar and tensor fluctuations using the Bessel function method and derive expressions for the scalar spectral tilt $n_s$, the tensor spectral tilt $n_T$,  and the tensor-to-scalar ratio $r$. Additionally, we explore the dual between the ultra-slow-roll inflation and slow-roll inflation. 
The organization of this paper is as follows: Section \ref{sec:bg} presents the background evolution and defines the slow-roll parameters. Section \ref{sec:result} calculates the power spectrum for scalar and tensor perturbations using the Bessel function method within the constant-roll condition.  Section \ref{sec:dual} explores  the dual behavior between ultra-slow-roll inflation and slow-roll inflation. Finally, section \ref{sec:con} draws the conclusions.

\section{Inflation with non-minimally derivative coupling}\label{sec:bg}
In this section, we derive the background evolution, define the slow-roll parameters, and  discuss the general formulae for the the scalar and tensor spectral for the inflationary models with non-minimally derivative coupling. 
\subsection{The background}
The action for the inflationary model with the non-minimal derivative coupling is
\begin{eqnarray} \label{action}
S=\frac{1}{2}\int\text{d}^{4} x \sqrt{-g}\bigg[M_{p l}^{2} R-g^{\mu \nu} \partial_{\mu} \phi \partial_{\nu} \phi \nonumber\\
+\frac{1}{M^{2}} G^{\mu \nu} \partial_{\mu} \phi \partial_{\nu} \phi-2 V(\phi)\biggr],
\end{eqnarray}
where $M_{pl} =1/\sqrt{8\pi G}$ is the reduced planck mass,  $R$ is the Ricci scalar, $\phi$ is the scalar field, $G_{\mu\nu}$ is the Einstein tensor, $V(\phi)$ is the potential, and $M^2$ is the coupling parameter with the dimension of energy. 
For the homogeneous and isotropic background with the flat  Friedmann–Robertson–Walker (FRW) metric,  the background equations are \cite{Yang:2015pga}
\begin{gather} \label{bkeq1}
H^{2}=\left(\frac{\dot{a}}{a}\right)^{2}=\frac{1}{3 M_{pl}^{2}}\left[\frac{\dot{\phi}^{2}}{2}(1+9 F)+V(\phi) \right],\\
\label{bkeq2}
\frac{{\rm d}}{{\rm d} t}\left[a^{3} \dot{\phi}\left(1+3 F)\right)\right]=-a^{3} \frac{{\rm d} V}{{\rm d} \phi},\\
\label{Ray1}
\dot{H}-\frac{\dot{\phi} \ddot{\phi}}{M^{2} M_{p l}^{2}} H=\frac{1}{2} \frac{\dot{\phi}^{2}}{M^{2} M_{p l}^{2}} \dot{H}-\frac{3}{2} \frac{\dot{\phi}^{2}}{M^{2} M_{p l}^{2}} H^{2}-\frac{1}{2} \frac{\dot{\phi}^{2}}{M_{p l}^{2}},
\end{gather}
where $F=H^2/M^2$ and an over-dot means the derivative with respect to the cosmic time $t$.  

In this paper, we define the slow-roll parameters as 
\begin{equation}\label{srp}
 \epsilon_{H}=-\frac{\dot{H}}{H^2},\quad \eta_{H}=-\frac{\ddot{H}}{2H\dot{H}}, \quad \eta_{\phi}=-\frac{\ddot{\phi}}{H\dot{\phi}}.
 \end{equation}
In the canonical inflation models, the slow-roll parameters satisfy $\eta_H =  \eta_\phi$, a condition not met in the inflation models with non-minimal derivative coupling. By using the slow-roll parameters \eqref{srp}, the background equation \eqref{Ray1} can be rewritten as
\begin{equation}\label{Ray2}
 - \epsilon_{H}+2\eta_{\phi}Q=-\epsilon_{H}Q-3Q-Q/F,
\end{equation}
where the definition of $Q$ is
\begin{equation}
\label{qdef1}
Q=\frac{\dot{\phi}^{2}}{2 M^{2} M_{pl}^{2}}.
\end{equation}
Solving the background equation \eqref{Ray2}, we get  
\begin{equation}
\label{A} 
Q=\frac{F \epsilon_{H}}{b+F \epsilon_{H}},
\end{equation}
where $b=1+3 F+2 \eta _{\phi}F$.

For later use, we also give the following relation,
\begin{equation}
\label{dotepsh}
\dot{\epsilon}_H=2H(\epsilon_H-\eta_H)\epsilon_H,    
\end{equation}
\begin{equation}
\label{dotfeq}   
\dot{F}=-2HF\epsilon_H,
\end{equation}
and
\begin{equation}\label{dah}
\frac{\text{d}}{\text{d} \tau}\left(\frac{1}{a H}\right)=-1+\epsilon_{H},
\end{equation}
where the conformal time $\tau$ is related to the cosmic time $t$ by $\text{d} t=a \text{d} \tau$.

\subsection{The perturbations}
In the uniform field gauge $\delta \phi(x,t) =0$,  the action for the primordial  scalar perturbation  $\zeta$ is   \cite{Germani:2010ux,Yang:2015pga, Germani:2011ua} 
\begin{equation}\label{action3}
    S_{\zeta^{2}}=\int \text{d} t \text{d}^{3} x M_{pl}^{2} a^{3}\left\{\frac{\Sigma}{\bar{H}^{2}} \dot{\zeta}^{2}-\frac{\theta_{s}}{a^{2}}\left(\partial_{i} \zeta\right)^{2}\right\},
\end{equation}
where 
\begin{eqnarray}\label{1}
 \bar{H}=\frac{H(1-3 Q )}{1-Q }, \quad\Sigma=Q M^{2}\left[1+\frac{3F(1+3 Q)}{(1-Q)}\right],
\end{eqnarray}
and 
\begin{equation}
    \theta_{s}=\frac{1}{a} \frac{\text{d}}{\text{d} t}\left[\frac{a}{\bar{H}}(1-Q)\right]-1-Q.
\end{equation}
By the variation principle, we can derive the equation of motion  governing  the primordial perturbation $\zeta$; and in the Fourier space, we get the   Mukhanov-Sasaki equation 
\begin{equation}\label{ms}
v_{k}^{\prime \prime}+\left(c_{s}^{2} k^{2}-\frac{z^{\prime \prime}}{z}\right) v_{k}=0,
\end{equation}
where the prime denotes the derivative with respect to the conformal time $\tau$,
the effective sound speed is $c_{s}^{2}=\bar{H}^{2} \theta_{s} / \Sigma$, the canonically normalized field is $v=z \zeta$, and 
\begin{equation}\label{z}
    z=a M_{p l} \frac{\sqrt{2 \Sigma}}{\bar{H}}. 
\end{equation}
By using the slow-roll parameters \eqref{srp} and equation \eqref{Ray2},  we have 
\begin{eqnarray}\label{cs}
c_{s}^{2}=&&1-\frac{2 \epsilon_{H}F(1+7 F+6 F \eta_{\phi})}{(1+3 F)^{2}+2 F(1+3 F)\eta_{\phi}+12 \epsilon_{H}F^{2}} \nonumber\\
&&-\frac{8\epsilon_{H}^{2}F^3}{b[b+3F(b+4F\epsilon_{H})]},
\end{eqnarray}
and 
 \begin{align}   
 \label{ztoz}
   \frac{z^{\prime \prime}}{z}
=&a^2H^{2}\bigg[2+3 \eta_{H}-4\epsilon_{H}-6 \eta_{\phi}-\eta_{\phi}(2 \epsilon_{H}-2 \eta_{H}+\eta_{\phi})\nonumber\\
&+\frac{\eta_{H}-\eta_{\phi}}{F}\biggr]+\frac{a^{2} H^{2}b\epsilon_{H}(3-2 \eta_{\phi}+3 \epsilon_{H}-2 \eta_{H})}{b+3 F(b+4 F \epsilon_{H})}\nonumber\\
&+\frac{8 a^{2} H^{2} F \epsilon_{H} m(b+F \epsilon_{H})^{2}}{b^{2}\left(b-2 F \epsilon_{H}\right) \left(b+4 F \epsilon_{H}\right)}\nonumber\\
&+\frac{48 a^{2} H^{2} F^{2} \epsilon_{H}^{2} \eta_{\phi}^{2}(b+F \epsilon_{H})}{b(b-2 F \epsilon_{H})^{2}}\nonumber\\
&+\frac{48 a^{2} H^{2} F^{3} \eta_{\phi}^{2}\epsilon_{H}^{3}(b+F\epsilon_{H})}{b^{2}(b+4 F \epsilon_{H})^2}\nonumber\\&-\frac{a^{2} H^{2} \epsilon_{H}^{2}[4 F\eta_{\phi} (b+F \epsilon_{H})+b(b+4F\epsilon_{H})]^{2}}{(b+4 F \epsilon_{H})^{2}[b+3 F(b+4 F \epsilon_{H})]^{2}}\nonumber\\
 &-\frac{4 a^{2} H^{2} F \epsilon_{H}(b+F \epsilon_{H})}{b+3 F(b+4 F \epsilon_{H})}\bigg[\frac{2 \eta_{\phi} \epsilon_{H}}{b-2 F \epsilon_{H}}
 +\frac{\eta_{H}-\eta_{\phi}}{F(b+4 F \epsilon_{H})}
 \nonumber\\
 &+\frac{3(\eta_{H}-\epsilon_{H}-2 \eta_{\phi})-2 \eta_{\phi}(2\epsilon_{H}-\eta_{\phi}-\eta_{H})}{b+4 F \epsilon_{H}}\nonumber\\
 &- \frac{4 F\epsilon_{H} \eta_{\phi}^{2}(b-2 F\epsilon_{H})}{b(b+4 F\epsilon_{H})^{2}}\nonumber\\
 &+\frac{8F\eta_{\phi}^{2} \epsilon_{H}(b+F\epsilon_{H})}{b(b-2 F\epsilon_{H})(b+4 F\epsilon_{H})}\biggr],
\end{align}
where
\begin{align}
m&=2\eta_{\phi}(\eta_{\phi}+\epsilon_{H})+4F \eta_{\phi}^{2} (\eta_{H} +\eta_{\phi})+(1+3 F)\times
\nonumber\\
&
\bigg[3(\eta_{H}-\epsilon_{H}-2 \eta_{\phi})
-2 \eta_{\phi}(2 \epsilon_{H}+\eta_{\phi}-2 \eta_{H})+\frac{\eta_{H}-\eta_{\phi}}{F}\biggr].
\end{align}
The definition of the power spectrum of the scalar perturbations is
\begin{equation}  \label{def:spectrumS}
\mathcal{P}_{\zeta}  =\frac{k^{3}}{2 \pi^{2}}\left|\zeta_{k}\right|^{2},
\end{equation} 
and the scalar spectral tilt $n_s$ is defined by \cite{Kinney:2005vj}
\begin{equation}\label{def:ns}
  n_s-1=\left.\frac{{\rm d}\ln \mathcal P_{\zeta}}{{\rm d} \ln k}\right|_{a H =\text{const}}.
\end{equation}

For the tensor perturbation,  the  quadratic action  is \cite{Yang:2015pga, Germani:2011ua}
\begin{equation}\label{actiont}
    S=\int \text{d}^{3} x \text{d} t \frac{M_{p l}^{2}a^{3}}{8} \left[(1-Q) \dot{\gamma}_{i j}^{2}-\frac{1}{a^{2}}(1+Q)(\partial_{l} \gamma_{i j})^{2}\right].
\end{equation}
The   Mukhanov-Sasaki equation   for tensor perturbation is
\begin{equation}\label{ms:gw}
    u_{k}^{s \prime \prime}+\left(c_{t}^{2} k^{2}-\frac{z_t''}{z_t}\right) u_{k}^{s}=0.
\end{equation}
Here,   $u^{s} = z_t \gamma^s$, the label $s =\{ +, \times\}$, denotes the polarization of the tensor perturbations,  and 
\begin{equation}
   \gamma_{i j}=\sum_{s=+, \times} \text{e}_{i j}^{s} \gamma^{s},
\end{equation}
where $\text{e}_{i j}^{s}$ is the polarizaiton tensor and  satisfies  $\sum_{i} \text{e}_{i i}^{s}=0$, $\sum_{i, j} \text{e}_{i j}^{s}\text{e}_{i j}^{s^{\prime}}=2 \delta_{s s^{\prime}}$.   The sound speed  in equation \eqref{ms:gw} is 
\begin{equation}
    c_{t}^{2}=\frac{1+Q}{1-Q},
\end{equation}
and the normalized parameter is 
\begin{equation}
    z_{t}=\frac{\sqrt{2}}{2} a M_{p l} \sqrt{1-Q}.
\end{equation}
By using the slow-roll parameters \eqref{srp},  equation \eqref{Ray2}, and the relation \eqref{A}, we obtain
\begin{equation}\label{ct}
    c_{t}^{2}=  1+\frac{2F\epsilon_{H}}{1+(3+2\eta_{\phi})F},
\end{equation}
and
\begin{align} 
\label{zt}
     \frac{z_{t}^{\prime \prime}}{z_{t}}=
     &a^{2} H^{2}\left[2-\epsilon_{H}-\frac{F\epsilon_{H}}{1+(3+2\eta_{\phi})F}\bigg(\frac{\eta_{H}-\eta_{\phi}}{F}\right.\nonumber\\
     &+3(\eta_{H}-2\eta_{\phi}-\epsilon_{H})
     -2\eta_{\phi}(\epsilon_{H}-\eta_{H})
     \bigg)\nonumber\\
     &-\frac{F^2\epsilon_{H}^2\eta_{\phi}^2}{[1+(3+2\eta_{\phi})F]^2}\biggr].
\end{align}
The definition of the power spectrum for the tensor perturbations is
\begin{equation}\label{def:Pt}
 \mathcal{P}_{T}   =\frac{k^{3}}{2 \pi^{2}}\left|\gamma_{i j}\right|^{2}=\frac{k^{3}}{\pi^{2}} \sum_{s=+, x}\left|\frac{u_{k}^{s}}{z_{t}}\right|^{2}.
\end{equation}


\section{The Constant-roll inflation} \label{sec:result}  
In addition to the slow-roll inflation scenario, where all the slow-roll parameters are small, the constant-roll inflation scenario, where one of the slow-roll parameters is held constant and need not be small, can also produce predictions consistent with observational data from  CMB.  In this section, we  study the constant-roll scenario where the slow-roll parameter $\eta_\phi$ is a constant.   

Combining equations \eqref{qdef1} and \eqref{A} and calculating $\dot{Q}$, we get 
\begin{equation}\label{dphi}
\begin{split}
   \dot{ \eta_{\phi}}=&3H( \eta_{\phi}- \eta_{H}+\epsilon_{H})+ \eta_{\phi}H(3\epsilon_{H}-2 \eta_{H}+2 \eta_{\phi})\\
   &-\frac{( \eta_{H}- \eta_{\phi})H}{F}.
\end{split}
\end{equation}
By using the condition that the slow-roll parameter $\eta_\phi$ is a constant,  we obtain
\begin{equation} \label{etahrel1}
\eta_H=\eta_\phi+\frac{3F\epsilon_H(\eta_\phi+1)}{1+(3+2\eta_\phi)F},   
\end{equation}
and relation \eqref{dotepsh} becomes, 
 \begin{equation}\label{depsilon}
    \dot{\epsilon}_H=2H\epsilon_H\left(\epsilon_H-\eta_\phi-\frac{3F\epsilon_H(\eta_\phi+1)}{1+(3+2\eta_\phi)F}\right).  
\end{equation}   
Because $\eta_\phi$ is a constant and it may not be small,
so $\dot{\epsilon}_H$ can still be the first order in perturbation.
 
In the GR limit, $F=0$, we recover the standard result $\eta_H=\eta_\phi$.
In the high friction limit, $F\gg 1$,
\begin{equation}
\label{etahrel2}   
\eta_H=\eta_\phi+\frac{3\epsilon_H(\eta_\phi+1)}{3+2\eta_\phi}.
\end{equation}
 Because the slow-roll parameter $\eta_\phi$ may not be small in the constant-roll inflation scenario, 
 the slow-roll parameter $\eta_H$ may not be small in general either.

For the constant-roll inflation with constant $\eta_\phi$, from equation \eqref{dah} and to the first order approximation of $\epsilon_{H}$, we get the relation of the conformal Hubble parameter $aH$ and the conformal time $\tau$ \cite{Yi:2017mxs},
\begin{equation}
\label{ah}
\frac{1}{aH}\approx\left(-1+\frac{\epsilon_{H}}{1+2\eta_{\phi}}\right)\tau.
\end{equation}
\subsection{Scalar Perturbations} 
To the first order of  $\epsilon_{H}$,  the relation \eqref{ztoz} becomes 
\begin{equation}
\begin{split}
\label{czz}
    \frac{z''}{z} \approx & a^{2} H^{2}\bigg[2-3 \eta_{\phi}+\eta_{\phi}^{2}+\frac{2-3 F}{1+3 F} \epsilon_{H}\\
    & -\eta_{\phi} \epsilon_{H}\left(\frac{3(1-F)}{1+3 F}  +\frac{4 F (1+6 F)(3 -4 \eta_{\phi})}{(1+3 F)[1+(3+2 \eta_{\phi}) F]}\right)\biggr].
 \end{split}     
 \end{equation} 
To solve the Mukhanov-Sasaki equation \eqref{ms},  
we write  
\begin{equation}\label{nu}
\frac{z''}{z}=\frac{\nu^2-1/4}{\tau^2}.
\end{equation} 
To the first order of  $\epsilon_{H}$,  
combining equations \eqref{ah}  and \eqref{czz},  we get
 \begin{eqnarray} 
 \label{nu1}
 \nu&=&\left|\frac{3}{2}-\eta_{\phi} \right|+\frac{\bar\nu \epsilon_{H}}{\left|3/2-\eta_{\phi}\right|},
\end{eqnarray}
where    
\begin{equation}
\begin{split}\label{nubar}
\bar\nu=&\frac{2-3 \eta_{\phi}+\eta_{\phi}^{2} }{1+2 \eta_{\phi}}+\frac{2-3 F}{2(1+3 F)} \\
&-\eta_{\phi}\left[\frac{3(1-F)}{2(1+3 F)} +\frac{2 F(1+6 F) (3-4 \eta_{\phi}) }{(1+3 F)[1+(3+2 \eta_{\phi}) F]}\right].
\end{split}
\end{equation}
Assuming that both $\nu$ and $c_s$ are constants, 
then the Mukhanov-Sasaki equation \eqref{ms} becomes the standard Bessel equation, 
and the solution can be expressed as the Hankel function of order $\nu$ \cite{Stewart:1993bc},
\begin{equation}\label{Hankel:func}
     v_{k}(\tau)=\sqrt{-\tau}\left[c_{1} H_{\nu}^{(1)}(-c_s k \tau)+c_{2} H_{\nu}^{(2)}(-c_s k \tau)\right] . 
\end{equation}
Choosing the  Bunch-Davies vacuum, 
at small scales, $-c_s k \tau \gg1$, we have the initial condition
\begin{equation}\label{vk0}
v_{k} \rightarrow \frac{1}{\sqrt{2 c_{s} k}} \text{e}^{\text{-i} c_{s} k \tau}. 
\end{equation}
Using the inital condition \eqref{vk0} and the asymptotic behavior of the Hankel function,  
we obtain the mode function outside the horizon
\begin{equation}\label{sol:vk}
    v_{k}=\text{e}^{\text{-i}(\nu-1 / 2) \frac{\pi}{2}} 2^{\nu-\frac{3}{2}} \frac{\Gamma(\nu)}{\Gamma(3 / 2)} \frac{1}{\sqrt{2 c_{s} k}}\left(-c_{s} k \tau\right)^{\frac{1}{2}-\nu}.
\end{equation}
Combining equation \eqref{z} and the solution \eqref{sol:vk}, from the definition of the  power spectrum for the scalar perturbation  \eqref{def:spectrumS}, we obtain  
\begin{eqnarray}
  \label{spectrumS}
\mathcal{P}_{\zeta}  
 \approx 2^{2 \nu-3}\left[\frac{\Gamma(\nu)}{\Gamma(3 / 2)}\right]^{2} \frac{H^{2}}{8 \pi^{2}M_{p l}^{2} c_{s} \theta_{s} } \nonumber\\
 \times\left(1-\frac{\epsilon_H}{1+2\eta_\phi}\right)^{2\nu-1}\left(\frac{c_{s}k}{aH}\right)^{3-2 \nu}.
\end{eqnarray} 
Substituting  equation \eqref{spectrumS} into the definition of the scalar spectral tilt  $n_s$, equation \eqref{def:ns}, we obtain 
\begin{equation}
\label{NS}
  n_s-1=  3-2\nu.
\end{equation}
If $|\eta_\phi|\ll 1$, to the first order of perturbation, we get \cite{Tsujikawa:2012mk,Yang:2015pga}
\begin{equation}
\label{srnsres}
n_s-1\approx 2\eta_\phi-\frac{2(2+3F)}{1+3F}\epsilon_H.   
\end{equation}

In the GR limit,  we  have
\begin{equation}
\nu=\left|\frac{3}{2}-\eta_{\phi}\right|+\frac{(6-5\eta_{\phi}-4\eta^2_{\phi})\epsilon_{H}}{\left|3-2\eta_{\phi}\right|(1+2\eta_{\phi})},
\end{equation}
and 
\begin{equation}
\label{grns1}
n_s-1=3-\left|3-2\eta_{\phi}\right|-\frac{2(6-5\eta_{\phi}-4\eta^2_{\phi})\epsilon_{H}}{\left|3-2\eta_{\phi}\right|(1+2\eta_{\phi})}.
\end{equation}

In the high friction limit,  we  have
\begin{equation}
    \nu=\left|\frac{3}{2}-\eta_{\phi}\right|+\frac{3(3-4\eta_{\phi})(1-3\eta_{\phi }-6\eta^2_{\phi})\epsilon_{H}}{\left|3-2\eta_{\phi}\right|(1+2\eta_{\phi})(3+2\eta_{\phi})},
\end{equation}
\begin{equation}
\begin{split}
n_s-1=&3-\left|3-2\eta_{\phi}\right|\\
&-\frac{6(3-4\eta_{\phi})(1-3\eta_{\phi }-6\eta^2_{\phi})\epsilon_{H}}{\left|3-2\eta_{\phi}\right|(1+2\eta_{\phi})(3+2\eta_{\phi})}.
\end{split}
\end{equation}

As discussed above, $\dot{\epsilon}_H$ is still the first order, 
so we will get a different expression for the scalar spectral tilt if we take the derivative of the amplitude of the scalar power spectrum at the horizon crossing $c_sk=aH$ with respect to the scale $k$. 
To show this point, we give the derivation below.

For the constant-roll inflation with large $\eta_\phi$, the scalar perturbations may continue evolving outside the horizon \cite{Motohashi:2014ppa}. If scalar perturbations are constant after the horizon crossing, 
$aH\geq c_s k$,  
to the first order of $\epsilon_{H}$, 
the  amplitude of the power spectrum for scalar perturbations outside the horizon is  
\begin{equation}
\begin{aligned}
 \label{Ps1}
\mathcal{A}_\zeta &= \mathcal{P}_{\zeta} |_{c_{s} k=a H} \\
&\approx  2^{|3-2 \eta_{\phi }|-3}\left[\frac{\Gamma(|3/2-\eta_\phi|)}{\Gamma(3 / 2)}\right]^{2} \frac{H^{2}}{8 \pi^{2}  M_{p l}^{2}\epsilon_{H}} \\
&\times \frac{1+3F+2F\eta_\phi}{1+3F}
 (1+D \epsilon_H), 
\end{aligned}
\end{equation}
where 
\begin{eqnarray}
\label{cdef1}
 D=&  &\frac{4\bar\nu}{|3-2 \eta_{\phi } |}\left(\ln2+\frac{\Gamma'(|3/2-\eta_\phi|)}{\Gamma(|3/2-\eta_\phi|)}\right)  
  -\frac{|3-2 \eta_{\phi } |-1}{1+2\eta_{\phi }}\nonumber\\
  &&\frac{18 F^{2} \eta_{\phi}}{(1+3 F)[1+(3+2\eta_{\phi})F]}.
\end{eqnarray}
Besides the definition \eqref{def:spectrumS}, 
if scalar perturbations are constants after the horizon crossing, 
the  scalar spectral tiltcan also be derived by using the  the  amplitude of the power spectrum for the scalar perturbation,
\begin{equation}\label{def:spectrumS2}
    n_s-1=\left.\frac{\text{d} \ln \mathcal{A}_{\zeta}}{\text{d} \ln k}\right|_{c_{s} k=a H}.
\end{equation}
To the first order of approximation of $\epsilon_H$, 
\begin{equation}
\label{dlnkeq}
\begin{split}
\frac{\text{d}\ln k}{H\text{d}t}&=(1-\epsilon_{H})-H^{-1}\frac{\dot{c_{s}}}{c_{s}}\\
    &\approx 1-\left(\frac{2 F\eta_{\phi}(1+7F+6F\eta_{\phi})  }{(1+3 F)[1+(3 +2  \eta_{\phi})F]}+1\right)\epsilon_{H}. \\
\end{split}
\end{equation}
Substituting  equations \eqref{Ps1} and \eqref{dlnkeq} into equation \eqref{def:spectrumS2},  
to the first order of $\epsilon_H$,  
we get
\begin{equation}
\begin{aligned}
\label{lnPb}
n_s&-1 \approx2 \eta_{\phi}-\frac{2(2+3F)\epsilon_{H} }{1+3 F}\\
&+2\eta_{\phi}\epsilon_{H}  \Bigg[ \frac{|3-2\eta_{\phi}|-1}{1+2\eta_{\phi}} \\
&-\frac{4\Bar{\nu} }{\left|3-2\eta_{\phi}\right|}\left(\ln2+\frac{\Gamma'(|3/2-\eta_\phi|)}{\Gamma(|3/2-\eta_\phi|)}\right)\\
&+\frac{(1+3F)(1+4F)+2F\eta_{\phi }(2+F+6F\eta_{\phi })}{(1+3 F)[1+(3+2  \eta_{\phi})F]}\biggr].    
\end{aligned}
\end{equation}
It is obvious this result is different from equation \eqref{NS} if $\eta_\phi$ is not small. 
If $|\eta_\phi|\ll 1$, then the result \eqref{lnPb} reduces to the result \eqref{srnsres}.
In the following discussion,
we use the familiar result \eqref{NS}.

\subsection{Tensor Perturbation}
For tensor perturbations, to solve equation \eqref{ms:gw}, 
we use the same method as that solving the scalar perturbation.  
To the first order of  $\epsilon_{H}$, we have
\begin{equation}
\label{zt2}
\begin{split}
\frac{z_{t}^{\prime \prime}}{z_{t}} &\approx  a^{2} H^{2}\left[2-\epsilon_{H}+\frac{F\epsilon_{H}\eta_{\phi}(3-2\eta_{\phi})}{1+(3+2\eta_{\phi})F}\right]\\
&= \frac{\mu^2-1/4}{\tau^2},
\end{split}
\end{equation}
where
\begin{equation}\label{mu2}
    \mu \approx \frac{3}{2}+\bar\mu \epsilon_{H},
\end{equation}
and
\begin{equation}
  \bar\mu=  \frac{(3-2 \eta_{\phi})[1+3F+F\eta_{\phi}(3+2 \eta_{\phi})]}{3(1+2 \eta_{\phi})[1+(3+2  \eta_{\phi})F]} .
\end{equation}

Like scalar perturbations, assuming that both $c_t$ and $\mu$ are constants,   
to the first order of $\epsilon_{H}$, 
we obtain the power spectrum for the tensor  perturbation,
\begin{equation}\label{Pt}
\begin{aligned}
  \mathcal{P}_{T}  & 
   \approx \frac{2^{2\mu}}{M_{pl}^{2}}\left(\frac{H}{2\pi}\right)^{2}\left[\frac{\Gamma(\mu)}{\Gamma(3 / 2)}\right]^{2}\left(1-\frac{\epsilon_H}{1+2\eta_{\phi}}\right)^{2\mu-1}\\ &\left(\frac{1+3F+2F\eta_{\phi}-2F\epsilon_{H}}{1+(3+2\eta_{\phi})F}\right)\left(\frac{c_tk}{aH}\right)^{(3-2\mu)}.
\end{aligned}
\end{equation}
Therefore, the tensor spectral tilt is
\begin{equation}
\label{NT}
\begin{split}
n_T&=\left.\frac{\text{d} \ln \mathcal P_{T}}{\text{d} \ln k}\right|_{a H =\text{const}}\\
&=-\frac{2(3-2 \eta_{\phi})[1+3F+F\eta_{\phi}(3+2 \eta_{\phi})]}{3(1+2 \eta_{\phi})[1+(3 +2  \eta_{\phi})F]}\epsilon_H.
\end{split}
\end{equation} 
If $|\eta_\phi|\ll 1$, then we get $n_T\approx-2\epsilon_H$.

In the GR limit, we get
\begin{equation}
\label{NT}
n_T=-\frac{2(3-2 \eta_{\phi})}{3(1+2 \eta_{\phi})}\epsilon_H.
\end{equation} 
In the high friction limit, we get
\begin{equation}
\label{NT}
n_T=-\frac{2(3-2 \eta_{\phi})[3+\eta_{\phi}(3+2 \eta_{\phi})]}{3(1+2 \eta_{\phi})(3+2 \eta_{\phi})}\epsilon_H.
\end{equation} 

To the first order of $\epsilon_{H}$, 
the amplitude of the power spectrum for 
tensor perturbations becomes
\begin{eqnarray}\label{pt2}
\mathcal{P}_{T}\bigg|_{c_{t} k=a H}
\approx\frac{8}{M_{pl}^2}\left(\frac{H}{2\pi}\right)^{2}\bigg[1-(A-B)\epsilon_{H}\biggr],
\end{eqnarray}

where
\begin{equation}
\begin{split}
A&=\frac{2[3+C(3-2\eta_{\phi})]}{3(1+2\eta_{\phi})},\\
B&=-\frac{2F[C\eta_{\phi}(3-2\eta_{\phi})+3]}{3[1+(3+2\eta_{\phi})F]},
\end{split}
\end{equation}
and  the constant $C=-2+\gamma+\ln{2}\approx-0.73 $. 

Combining the scalar power spectrum \eqref{Ps1} and the tensor power spectrum \eqref{pt2},
we obtain the  tensor-to-scalar ratio 
\begin{equation} \label{r1}
\begin{aligned}
r &=\frac{\mathcal{P}_T|_{c_{t} k=a H}}{\mathcal{P}_\zeta|_{c_{s} k=a H}} \\
 &=\frac{2^{7-|3-2\eta_{\phi}|} [\Gamma(3/2)]^2 (1+3F) }{\left[\Gamma(|3/2-\eta_{\phi}|)\right]^2[1+(3+2\eta_{\phi})F]} \epsilon_H.  
\end{aligned}
\end{equation}
It is apparent that the consistency relation $r=-8n_T$ does not hold for the constant-roll inflation if $|\eta_\phi|$ is not small.
If $|\eta_\phi|\ll 1$, then 
\begin{equation}
\label{r4}
r\approx 16 \epsilon_H,
\end{equation}
the consistency relation still holds.

In the GR limit, we get 
\begin{equation}
\label{r2}
r=\frac{2^{7-|3-2\eta_{\phi}|}[\Gamma(3/2)]^2}{\left[\Gamma(|3/2-\eta_{\phi}|)\right]^2}\epsilon_H.
\end{equation}
In the high friction limit, we get
\begin{equation}
\label{r3}
r=\frac{3 [\Gamma(3/2)]^2 2^{7-|3-2\eta_{\phi}|} }{\left[\Gamma(|3/2-\eta_{\phi}|)\right]^2(3+2\eta_{\phi})} \epsilon_H.
\end{equation}

\section{The duality in constant-roll inflation}\label{sec:dual}

In the GR case, from equation \eqref{grns1}, 
we see that if we neglect the contribution from $\epsilon_H$,
then we have the same scalar spectral tilt $n_s-1\approx 2\alpha$
when we replace $\eta_\phi=\alpha$ by $\eta_\phi=3-\alpha$ with $|\alpha|<3/2$.
From equation \eqref{r2}, we see that the duality between $\eta_\phi=\alpha$ and $\eta_\phi=3-\alpha$ also exists for the tensor-to-scalar ratio $r$.
The duality is also referred to as the duality between constant slow-roll and ultra-slow-roll inflation   and explained as the two branches of the solution of the  inflation field from the same potential \cite{Morse:2018kda} 
although the behaviors of background and perturbations are different for the constant slow-roll and ultra-slow-roll inflationary models if $\epsilon_H$ is not negligible \cite{Gao:2019sbz}. 
In this section, we discuss this duality behavior, 
aiming to understand the connection between the slow-roll and ultra-slow-roll scenarios within the framework of inflationary models with non-minimally derivative coupling.

Combining equations \eqref{nu1} and \eqref{NS}, and neglecting the contribution from $\epsilon_H$, we get
\begin{equation}
\label{dualns1}
n_s-1= 3-|3-2\eta_\phi|.
\end{equation}
It is obvious that the duality between $\eta_\phi=\alpha$ and $\eta_\phi=3-\alpha$  holds, and
\begin{equation}
 n_s-1  \approx 2 \alpha,
\end{equation}
if $|\alpha|<3/2$.
Now we discuss whether the duality holds for the tensor-to-scalar ratio $r$.
For the constant slow-roll inflation with $\eta_\phi=\alpha$, from equation \eqref{r1}, we get
\begin{equation}
\label{r6}
r=\frac{2^{4+2\alpha} [\Gamma(3/2)]^2 (1+3F) }{\left[\Gamma(3/2-\alpha)\right]^2[1+(3+2\alpha)F]} \epsilon_H.
\end{equation}
For the ultra-slow-roll inflation with $\eta_\phi =3 -\alpha$,  the tensor-to-scalar ratio becomes
\begin{equation}
\label{r7}
r=\frac{2^{4+2\alpha} [\Gamma(3/2)]^2 (1+3F) }{\left[\Gamma(3/2-\alpha)\right]^2[1+(9-2\alpha)F]} \epsilon_H.
\end{equation}
Comparing equations \eqref{r6} and \eqref{r7}, we find that they are the same only if $F=0$,
i.e., the duality exists only in the GR case, and there is no such duality in inflationary models with non-minimally derivative coupling.

When the slow-roll parameter $\epsilon_H$ is extremely small and can be neglected, 
the scalar spectral tilts obtained from both the slow-roll inflation and the ultra-slow-roll inflation coincide. 
In the limit where $F=0$ (GR limit), the tensor-to-scalar ratio \eqref{r6} derived from slow-roll inflation matches equation \eqref{r7} obtained from ultra-slow-roll inflation, 
thus the duality recovers in the GR limit.
However, in more general case, the tensor-to-scalar ratios $r$  from slow-roll inflation and ultra-slow-roll inflation are different. 
Consequently, even for the scalar spectral tilt $n_s$ and the tensor-to-scalar ratio $r$, 
the duality behavior between slow-roll and ultra-slow-roll inflation models with non-minimally derivative coupling is not observed. 
Therefore, the duality of $n_s$ and $r$ is not a universal feature in the constant-roll inflation models. 

\section{conclusion} 
\label{sec:con}

We study the constant-roll inflation model with the kinetic term non-minimally derivative coupling to the Einstein tensor. 
For the constant-roll condition, we choose the slow-roll parameter $\eta_\phi  =-\ddot{\phi}/( H\dot{\phi})$ to be a constant. 
With this constant-roll condition, we calculate the power spectra for the scalar and tensor perturbations and subsequently derive the scalar spectral tilt $n_s$, 
tensor spectral tilt $n_T$, and tensor-to-scalar ratio $r$. 
Due to the effect of large $\eta_\phi$, 
$\dot{\epsilon}_H$ remains to be the first order,
the expressions for $n_s$ are different with different ordering of taking the derivative of the scalar power spectrum with respect to the scale $k$ and the horizon crossing condition $c_sk=aH$ in the constant-roll inflation,
and the consistency relation between the tensor-to-scalar ratio and tensor spectral tilt, $r=-8n_T$, does not hold.
Between the slow-roll inflation with $\eta_\phi =\alpha$ being small and the ultra-slow-roll inflation with $\eta_\phi = 3-\alpha$, 
although the duality for the scalar spectral tilt still holds if we neglect the contribution of $\epsilon_H$ to $n_s$,
but the results for the tensor-to-scalar ratio $r$ are different. 
Thus, the duality of $n_s$ and $r$ is not a universal feature in inflation with non-minimally derivative coupling.
 
The inflation model with non-minimally derivative coupling is important because the new Higgs inflation model saves the Higgs inflation from unitarity bound violations. 
This investigation into constant-roll inflation presents a different perspective from the typical slow-roll inflation scenario, thereby enhancing our understanding of inflation models with non-minimally derivative coupling.  Scalar-induced gravitational waves usually originate from inflation models featuring a transitional ultra-slow-roll phase. Therefore, this investigation into the ultra-slow-roll condition is also helpful for the issue of scalar-induced gravitational waves in the inflation model with non-minimally derivative coupling.

\begin{acknowledgments}
JL is supported by the Hainan Provincial Natural Science Foundation of China under Grant No.121MS033. YG  is partially supported by  the National Key Research and Development Program of China under Grant No. 2020YFC2201504. ZY is supported by the National Natural Science Foundation of China under Grant No. 12205015.
\end{acknowledgments}


\end{document}